\documentclass[preprint,floatfix,amsmath,amssymb,aps,prb,showpacs]{revtex4}
\usepackage{epsfig}
\usepackage{amsmath}

\usepackage{soul}
\RequirePackage{color}
\definecolor{MyDarkGreen}{rgb}{0.02,0.60,0.06}

\def\la{\langle}
\def\ra{\rangle}

\begin{document}

\title{Monte-Carlo study of anisotropic scaling generated by disorder}

\author{O.Vasilyev}
\affiliation{Max-Planck-Institut f\"ur Intelligente Systeme,
Heisenbergstra{\ss}e 3, D--70569 Stuttgart, Germany}
\affiliation{ IV. Institut f{\"u}r Theoretische Physik,
  Universit{\"a}t Stuttgart,  Pfaffenwaldring 57, D--70569 Stuttgart, Germany}
\author{B. Berche}
\affiliation{Statistical Physics Group, Institut Jean Lamour, UMR CNRS 7198, Universit\'e de Lorraine, B.P. 70239, 54506
Vand{\oe}uvre l\`es Nancy Cedex, France}
\author{M. Dudka}
\affiliation{Institute  for Condensed   Matter  Physics,  National  Acad.  Sci.   of
Ukraine, UA--79011 Lviv, Ukraine}

\author{Yu. Holovatch}
\affiliation{Institute  for Condensed   Matter  Physics,  National  Acad.  Sci.   of
Ukraine, UA--79011 Lviv, Ukraine}

\date{\today}

\begin{abstract}
We analyze the critical properties of the three-dimensional Ising model with linear parallel extended defects. Such a form of disorder produces two distinct correlation lengths, a parallel correlation length $\xi_\parallel$ in the direction along  defects, and a perpendicular correlation length $\xi_\perp$ in the direction perpendicular to the lines. Both $\xi_\parallel$ and $\xi_\perp$ diverge algebraically in the vicinity of the critical point, but the corresponding critical exponents $\nu_\parallel$ and $\nu_\perp$ take different values. 
This property is specific for  anisotropic scaling and the ratio $\nu_\parallel/\nu_\perp$ defines the anisotropy exponent $\theta$. Estimates  of quantitative characteristics of the critical behaviour for such systems  were only obtained up to now within the renormalization group approach.  We report a study of the anisotropic scaling in this system via Monte Carlo simulation of the three-dimensional system with Ising spins and non-magnetic impurities  arranged into randomly distributed parallel lines. Several independent estimates for the anisotropy exponent $\theta$ of the system are obtained, 
as well as  an estimate of  the susceptibility exponent $\gamma$. Our results corroborate the renormalization group predictions obtained earlier.
\end{abstract}


\maketitle

\section{Introduction}\label{I}

The effect of structural disorder on  criticality remains one of the most attractive issues in condensed matter physics.~\cite{disorder}  Realistic systems always contain some imperfections of their structure. Thus  the question  of how disorder does influence 
the critical properties of a system deserves considerable interest. Obviously the modifications introduced depend on the amount of disorder 
as well as on the type of disorder. Quenched disorder is usually  studied in the form of  dilution (random site~\cite{dilution-site} or random bond~\cite{dilution-bond} systems), or
as a random field,~\cite{rfields} random connectivity\cite{rconnectivity} or  
anisotropy.~\cite{ranisotropy,Dudka05} 

In our study we will focus  on the effect of weak disorder in  the form of random sites.
In the simplest case, such disorder may be considered as uncorrelated, randomly distributed  point-like defects. The  relevance of point-like disorder for critical behaviour is predicted by the Harris criterion:~\cite{Harris74} it changes the universality class of $d$-dimensional system if the heat capacity critical exponent $\alpha_{\rm pure}$ of the corresponding homogenous (pure) system is positive. Since in $d=3$  the  pure Ising  model has $\alpha_{\rm pure}>0$,  weak point-like disorder there leads to a new critical behaviour. Results of  analytical, numerical, and experimental studies of this celebrated system 
are reviewed in Ref.~\onlinecite{dilution-site}.

Many real  systems  contain   more complex forms of disorder, for instance dislocations, disordered layers, grain boundaries, cavities or other extended defects. To take this into account Weinrib and Halperin  have  proposed a  model,\cite{Weinrib83} in which defects are correlated with a correlation function decaying with distance $x$ according to a power law: $g(x)\sim x^{-a}$. There is possible interpretation for integer value of $a$: the case $a=d-1$ ($a=d-2$) describes lines (planes) of random orientation, while  $a=d$ corresponds to above-mentioned uncorrelated defects. The critical properties of such systems have been extensively studied by 
the renormalization group (RG) approach~\cite{Weinrib83,Korutcheva,Prudnikov_RG,Holovatch_RG} as
well as through Monte Carlo (MC) simulations~\cite{Ballesteros99,Prudnikov_MC,Ivaneyko08}.

Another possible implementation of extended defects was proposed by 
Dorogovtsev~\cite{Dorogovtsev80} within  the model of a $d$-dimensional
 spin system with quenched random impurities that are strongly correlated in $\varepsilon_d$
dimensions and randomly distributed over the remaining
$d-\varepsilon_d$ dimensions. On the contrary of the model of Weinrib and Halperin which is isotropic,~\cite{Weinrib83} Dorogovtsev's model  describes a system which is expected to
behave differently along the directions ``parallel" to the
$\varepsilon_d$-dimensional impurity and in  the ``perpendicular"
hyper-planes. The case $\varepsilon_d=0$ corresponds to point-like
defects, and extended parallel linear (planar) defects are respectively given by $\varepsilon_d=1(2)$. Generalization of
$\varepsilon_d$ to nonnegative real numbers  may be interpreted as
an effective fractal dimension of a complex random defect system. However,
the relation of analytically continuation to non-integer Euclidean
dimension and a fractal dimension is not straightforward.\cite{fractal_dim} 
Critical behaviour of these systems was extensively studied by RG methods with the help of double $\epsilon=4-d,\,\epsilon_d$-expansions \cite{Dorogovtsev80,Boyanovsky82} as well as applying resummation technique
 to the RG asymptotic series directly at $d=3$ and fixed $\epsilon_d$.\cite{Lawrie84,Blavatska02,Blavatska03,Blavatska05}

There exist also investigations in the mean-field approximation\cite{Berche98} as well as MC simulations which
have some connections with Dorogovtsev's model with Ising spins and
extended defects with $\epsilon_d=2$.\cite{Lee92,Vojta}  In these latter studies, disorder was modeled by random bonds between planes of spins.
To our
best knowledge, the model with parallel linear extended defects in dimension $d=3$ has not been 
studied numerically so far, although the existing analytical predictions are accurate enough 
to challenge MC  verification. Therefore   we present such  MC study in our paper. The aim of this paper is therefore precisely to report such an extended numerical study.

The paper is organized as follows. In the next section we briefly present the analytical  predictions for scaling of three dimensional (3d) system with parallel extended defects as well as the RG estimates for the critical exponents of the Ising system with randomly distributed parallel linear defects. In Section~\ref{III} we remind the essentials of anisotropic finite size scaling. The formulation of our model, definition of the observables as well as details of the simulation are listed in Section~\ref{IV}. We eventually present the results of simulations in Section~\ref{V} and Section~\ref{VI} summarizes our study.

\section{Analytical results for a spin model with extended defects}\label{II}

The critical behaviour of the model under consideration is
described by the following effective Hamiltonian:
\begin{eqnarray} \label{eff_ham}
{\cal H}&=&\int d^dx\Big[\frac{1}{2}\big(\mu_0^2+V(x) \vec{\phi}^{\ \!2}(x)+
(\nabla_{\perp} \vec{\phi}(x))^2  \nonumber\\&&+\alpha_0({\nabla}_{||}
{\phi}(x))^2\big)+\frac{u_0}{4!}(\vec{\phi}^{\ \!2}(x))^2\Big].
\end{eqnarray}
Here, $\vec{\phi}$ is an $m$-component vector field: $\vec{\phi}=\{
\phi^{1}\cdots\phi^{m}\}$, $\mu_0$ and $u_0$ are the bare mass and
the coupling of the magnetic model, $\alpha_0$ is the bare
anisotropy constant, and $V(x)$ represents the impurity potential, which is assumed to be Gaussian distributed with zero mean and correlator:
\begin{eqnarray} \label{disorder_corr}
\overline {V(x)V(x')}
=&-v_0\delta^{d-\varepsilon_d}(x_{\perp}-{{x'}\!_{\perp}}). \,\,\,\,\,\,
\label{corr}
\end{eqnarray}
Here, overline stands for the average over
the potential distribution, (-$v_0$) is a positive constant
proportional to both the concentration of impurities and the
strength of their potential.  The impurities are envisaged as
$\varepsilon_d$-dimensional objects, each extending throughout the
system along the coordinate directions symbolized as $x_{||}$,
whereas in the remaining $d-\varepsilon_d$ dimensions they are
randomly distributed. Operators ${\nabla}_{\perp}$ and
${\nabla}_{||}$ mean differentiation in the coordinates $x_{\perp}$
and $x_{||}$, correspondingly. One assumes that the linear size of the
defects is much larger than the spin-correlation length and also
larger than the linear separation between any pair of defects. This
assumption is valid for defect concentrations well below the
percolation threshold.

The static critical behavior\cite{Dorogovtsev80,Boyanovsky82,Lawrie84,Blavatska02,Blavatska03,Blavatska05}  of this model  
as well as critical dynamics near equilibrium \cite{Prudnikov83,Lawrie84,Blavatska05,Blavatska06} were examined by means of the
RG method.
A double expansion in both $\varepsilon,\, \varepsilon_d$ was
suggested and RG functions were calculated to the first order.\cite{Dorogovtsev80} 
These results were consistent with a crossover to a new universality class in
the presence of extended defects. These calculations were
extended to the second order in Ref.~\onlinecite{Boyanovsky82}. Here, it
was argued that the Harris criterion is modified in the presence of
extended impurities: randomness is relevant, if
\begin{equation} \label{3}
\varepsilon_d > d - \frac{2}{\nu_{\rm pure}},
\end{equation}
where $\nu_{\rm pure}$ is the correlation length critical exponent
of the pure system. For point-like defects ($\varepsilon_d=0$) 
Eq.~(\ref{3}) reproduces the usual Harris criterion. Resummation technique
applied to the RG asymptotic series
\cite{Lawrie84,Blavatska02,Blavatska03,Blavatska05} lead to reliable
estimates of the critical exponents for this model. Furthermore,
different scenarios for the effective critical  behaviour were
discussed.\cite{Blavatska05}   Within RG approach, the influence of  cubic anisotropy of the order parameter\cite{Yamazaki} and the effect of replica symmetry  
breaking on the disorder average\cite{rsb} were studied. Reports of short time critical dynamics in systems with extended defects are also available in the 
literature.\cite{Fedorenko}
For completeness, we also mention here
 several papers more, where models with more complex  
 forms of disorder, including extended defects as a particular case,  
 were analyzed.\cite{deCesare94,Korzhenevski96}

The model described by the effective Hamiltonian~(\ref{eff_ham}) has rich
scaling behavior. As it was  already mentioned, such a system is no longer isotropic.
 Due to the spatial
anisotropy, two correlation lengths exist, one perpendicular and one
parallel to the extended impurities direction: ($\xi_{\perp}$ and
$\xi_{||}$). As the critical temperature $T_c$ is approached, their
divergences are characterized by corresponding critical exponents
$\nu_{\perp}$, $\nu_{||}$:
\begin {equation} \label{xi}
\xi_{\perp}\sim |t|^{-\nu_{\perp}},\phantom{5555555}
\xi_{||}\sim |t|^{-\nu_{||}},
\end{equation}
where $t$ is the reduced distance to the critical temperature
$t=(T-T_c)/T_c$.  The correlation of the order parameter
fluctuations in two different points acquires an orientational dependence.\cite{Dorogovtsev80} Thus, the critical
exponents $\eta_{\perp}$ and $\eta_{||}$, that characterize the
behavior of the correlation function in the directions,
perpendicular and parallel to the extended defects, must be
distinguished. Spatial anisotropy also modifies the critical dynamics near equilibrium producing two dynamical 
exponents $z_{||}$ and $z_{\perp}$.\cite{Prudnikov83} On the other hand, as far as the interaction of all
order parameter components with defects is the same, the system
susceptibility is isotropic \cite{Dorogovtsev80} and can be
expressed by the pair correlation function:\cite{Yamazaki}
\begin{eqnarray}
\chi(k_{\perp},k_{||},t)&=&
|t|^{-\gamma}g\left(\frac{k_{\perp}}{|t|^{\nu_{\perp}}},
\frac{k_{||}}{|t|^{\nu_{||}}},\pm 1\right)   \nonumber\\&=&  \left\{
 \begin{array}{ll}
k_{\perp}^{\eta_{\perp}-2}
g\left(1,\frac{k_{||}}{k_{\perp}^{\nu_{||}/\nu_{\perp}}},
\frac{|t|}{k_{\perp}^{1/\nu_{\perp}}}\right),& \\ k_{||}^{\eta_{||}-2}
g\left(\frac{k_{\perp}}{k_{||}^{\nu_{\perp}/\nu_{||}}},1,
\frac{|t|}{k_{||}^{1/\nu_{||}}}\right).& \end{array}\right. \label{chi1}
\end{eqnarray}
In (\ref{chi1}), $k_{||}$, $k_{\perp}$ are the components of the
momenta along $\varepsilon_d$ and $d-\varepsilon_d$ directions,
respectively, $\gamma$ is the magnetic susceptibility critical
exponent  and $g$ is the scaling function. The following scaling relations hold:\cite{Boyanovsky82,Lawrie84,Yamazaki}
\begin{equation}
\gamma=(2-\eta_{\perp})\nu_{\perp}=(2-\eta_{||})\nu_{||}.
\label{scaling1}
\end{equation}
The critical exponent $\alpha$ of the specific heat is related to
$\nu_{\perp}$, $\nu_{||}$ by the hyperscaling relation that differs
from the ordinary one:\cite{Dorogovtsev80}
\begin{equation}
\alpha=2-(d-\varepsilon_d)\nu_{\perp}-\varepsilon_d\nu_{||}.
\label{scaling2}
\end{equation}
All the other scaling relations are of the standard form. This
implies that one should calculate at least three independent
static exponents ({\it e.g.}, $\nu_{||},\nu_{\perp},\gamma$) instead of two, as
in the standard case, to find the others static exponents by scaling relations.

We are mostly interested in the case of Ising spins $m=1$ and linear parallel extended defects $\epsilon_d=1$. For this case we list below the numerical estimates of critical 
exponents obtained within RG approaches. These estimates based on ${\sqrt \varepsilon}$ expansions  are the following:\cite{Lawrie84}
\begin{equation}\label{epsilon}
\nu_{\perp}=0.67,\  \nu_{||}=0.84,\  \gamma=1.34,\  z_{\perp}=2.67,\  z_{||}=2.22.
\end{equation}
More reliable estimates are obtained with help of resummation of two-loop RG functions for static 
exponents:\cite{Blavatska03}
\begin{equation}\label{res_stat}
\nu_{\perp}=0.750,\  \nu_{||}=0.880,\  \gamma=1.483,
\end{equation}
  and for dynamic exponents:\cite{Blavatska05}
\begin{equation}\label{res_dyn}
 z_{\perp}=2.418,\  z_{||}=2.217.
\end{equation}
It is desirable to check
these results with MC simulations. In the following section we show how finite size scaling may be used to  extract quantitative characteristics of the critical behaviour from MC data obtained for finite systems.

\section{Anisotropic finite size scaling} \label{III}
Systems studied by MC simulations have limited sizes. 
Therefore finite size scaling (FSS) \cite{Fisher71,Fisher72,Barber83,Privman90,Privman94} 
plays a key role to extract the critical exponents of thermodynamical functions from  
MC data.  According to  FSS theory, in the vicinity of critical point $t\to 0$ 
the order parameter $M$ and the susceptibility $\chi$  of isotropic systems scale 
with the linear size  $L$ as:
\begin{eqnarray}\label{fss1}
{ M(t\to 0)}&=&L^{-\beta/\nu}{\mathcal M}(L),\\
{\chi(t\to 0)}&=&L^{\gamma/\nu}{\tilde \chi}(L)\label{fss2},
\end{eqnarray}
where $\mathcal M$, $\tilde \chi$ are magnetization and  
susceptibility scaling functions.
A quantity of special interest often used within MC simulation, namely of the fourth-order 
Binder's cumulant,\cite{Binder81} obeys the following behavior:
\begin{equation}\label{bind}
U_4(t,L)=1-\frac{\la M^4\ra}{3 \la M^2\ra^2}={\tilde U}_4{(tL^{1/{\nu}})},
\end{equation}
where $\la M^2 \ra$, $\la M^4 \ra$ 
are second and fourth moments of the distribution of order parameters,
${\tilde U}_4$ is the scaling function. 
In isotropic models all curves of the temperature dependence of 
$U_4$ for different system sizes intersect in one point, 
indicating the location of the critical temperature. 
It is true even if one keeps different aspect ratios for the simulation box: 
the curves for different sizes at a given generalized aspect ratio all intersect in the thermodynamic 
limit at the critical temperature.\cite{Kaneda99}

There exist different systems that manifest strong anisotropic scaling at criticality, 
{\em i.e.} the critical exponents for their correlation lengths differ in  different directions. Among them we mention as examples the behaviour of the next nearest neighbor Ising  model at the Lifshitz point,\cite{Lifshitz} non-equilibrium phase transition in  driven diffusive systems,\cite{driven} dipolar in-plane Ising model,\cite{dipolar} driven Ising model with friction,\cite{friction} Ising model under shear,\cite{shear,shear2} interface 
localization-delocalization problem.\cite{interface}  A common feature of the above mentioned systems is the presence a single anisotropy axis which results in the existence  of  two characteristic length scales: $L_{||}$ along the anisotropy axis and $L_{\perp}$ in the perpendicular directions.
According to the generalization of the FSS concept for systems with two characteristic length scales,\cite{Binder89} the properties of the thermodynamical functions depend on the ``generalized aspect ratio'' $L_{||}/L_{\perp}^{\theta}$, with  an anisotropy exponent $\theta=\nu_{||}/\nu_{\perp}$. Therefore relations (\ref{fss1}), (\ref{fss2})  are modified:
\begin{eqnarray}\label{ordpar}
{ M(t\to 0)}&=&L_{||}^{-\beta/\nu_{||}}{\mathcal M}(L_{||}/L_{\perp}^{\theta}),\\ 
\label{susc}
{\chi(t\to 0)}&=&L_{||}^{\gamma/\nu_{||}}{\tilde \chi}(L_{||}/L_{\perp}^{\theta}). 
\end{eqnarray}
Note that (\ref{ordpar}), (\ref{susc}) can be equivalently represented as:
\begin{eqnarray}
{ M(t\to 0)}&=&L_{\perp}^{-\beta/\nu_{\perp}}{\mathcal M}(L_{\perp}/L_{||}^{1/\theta}),\\
{ \chi(t\to 0)}&=&L_{\perp}^{\gamma/\nu_{\perp}}{\tilde \chi}(L_{\perp}/L_{||}^{1/\theta}).
\end{eqnarray}
Anisotropic finite size scaling  for  Binder's cumulant at the critical point predicts
\begin{equation}\label{binder_anis}
U_4(t\to 0)={\tilde U}_4(L_{||}/L_{\perp}^{\theta}).
\end{equation}

Therefore, one has to keep generalized aspect ratio $L_{||}/L_{\perp}^{\theta}$  fixed while performing MC simulations to extract critical exponents.
One can use the values of the exponents (\ref{epsilon})-(\ref{res_dyn}) to evaluate the anisotropy exponent $\theta$ for the model under consideration. 
Using results of ${\sqrt \varepsilon}$ expansions  (\ref{epsilon}) we can estimate $\theta=\nu_{||}/\nu_{\perp}\simeq 1.25$. Taking into account that  $\nu_{||}/\nu_{\perp}=z_{\perp}/z_{||}$ \cite{Lawrie84} we can get the second estimate of the ${\sqrt \varepsilon}$ expansions: $\theta=z_{\perp}/z_{||}\simeq 1.20$.   Next estimates can be obtained with  help  of the results of resummation of two-loop RG functions: $\theta=\nu_{||}/\nu_{\perp}=1.173$ from (\ref{res_stat}) and $\theta=z_{\perp}/z_{||}=1.091$ from (\ref{res_dyn}). All estimates give $\theta>1$. This reflects theÊ physical situation: in parallel directions, the fluctuations are stronger because they are not limited by defects and the correlation length in this direction  diverges more sharply.

In order to obtain an estimate of $\theta$ directly from the MC simulations, independently of the RG predictions, we can make use of the relation that characterizes systems possessing anisotropic scaling when $L_{\perp}\gg L_{||}$  at 
$t\to 0$:\cite{Henkel01,dipolar}
 \begin{equation}\label{relat}
 \xi_{\perp}(L_{||})\sim L_{||}^{1/\theta}.
 \end{equation}
Such a relation was successfully applied  for the analysis of the second order phase transitions in Ising models with frictions \cite{friction} and under shear.\cite{shear2}

In the following section, we use the formulae (\ref{ordpar})-(\ref{relat}) to perform the analysis of the results of  MC simulations. 

\section{Details of the simulations}\label{IV}

We consider the 3d Ising model on a simple cubic lattice
of  $L_{x}\times L_{y} \times L_{z}=V$ sites. Each site $i=(x,y,z)$,
$1 \le x \le L_{x}$, $1 \le y \le L_{y}$, $1 \le z \le L_{z}$, is characterized by the occupation number $c_{i}=\{0,1\}$ with $c_{i}=1$ if the site $i$ is occupied by a spin $s_{i}=\pm 1$, while  $c_{i}=0$ corresponds to a non-magnetic site $i$. Nearest neighbour spins interact ferromagnetically with constant exchange  $J>0$.
Explicitly, the Hamiltonian of the model may be written
\begin{equation}
{\mathcal H}= -J\sum \limits_{\la i,j \ra}
c_{i}c_{j}s_{i}s_{j},
\end{equation}
where the sum $\la i,j \ra$
is taken over the pairs of  nearest-neighbor spins. We use periodic boundary conditions. 

\begin{figure}
\includegraphics[width=0.25\textwidth,height=0.25\textwidth]{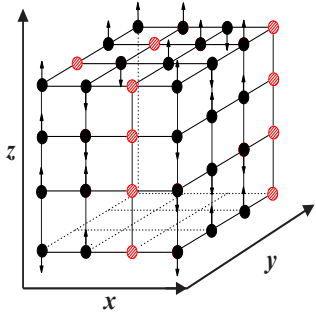}
\caption{
(Color online) An example of the lattice of size $4 \times 4 \times 4$
with Ising and (dark circles)
 non-magnetic impurities
 (light circles, red online) collected into $n_{\mathrm{imp}}=4$ lines.
}
\label{fig1}
\end{figure}

The non-magnetic sites ($c_{i}=0$) in our model have to be arranged into parallel lines and oriented along a given axis, say the  $z$ axis. The length of 
these lines coincides with the system size in $z$ direction, $L_{z}$.
 Fig.~\ref{fig1} shows one possible configuration of such lines.

We generate the impurity distribution for a given impurity concentration $p$
for a lattice of size  $V$
in the following way. We compute the integer number
$n_{\mathrm{\mathrm{imp}}}= {\rm int}( p L_{x} L_{y}) $
of impurity lines of length $L_{z}$.
We distribute randomly intersections of 
these lines with the $x-y$ plane,
as shown in Fig.~\ref{fig1}.

We compute powers  of the magnetization ($k=1,2,4$)
\begin{equation}
 M^{k} = \Bigl(\frac{1}{V} \sum \limits_{\{ i\}}
c_{i}s_{i} \Bigr)^{k},
\end{equation}
where $\{i \}$
means the lattice summation over all sites.
From powers of the magnetization we construct
the magnetic susceptibility and Binder cumulant in the following way
\begin{equation}
\chi=\beta V \overline{\bigl(\la M^{2} \ra -
\la | M| \ra^{2} \bigr)},
\end{equation}
\begin{equation}
\label{eq:u4}
U_{4}=1-\overline{\left(\frac{ \la M^{4}\ra}{3\la M^{2}\ra^2} \right)}.
\end{equation}
Here at first step we perform the thermal averaging 
of the power of magnetization for 
each impurity realization (denoted by angle brackets, {\it e.g.} $\la M^{2}\ra$).
Then we average these results over different impurity realizations. 
The averaging over impurity realizations is denoted by the overline.
In our simulation we fix $J/k_B=1$,  
in which case  $\beta=1/T$ is the inverse temperature.

We  compute also the correlation lengths for directions parallel to impurity lines  $\xi_{||}$ and perpendicular to them  $\xi_{\perp}$
 with  the help of the Fourier transform in a similar way as it was done for isotropic systems.\cite{Cooper89}
Let us introduce
$s({\bf k})=\sum \limits_{j} {\mathrm e}^{i {\bf k}\cdot j}c_j s_{j}$
where the wave vector is ${\bf k}=(k_{x},k_{y},k_{z})$
and we denote $\la \chi_{0} \ra= \la |s(0,0,0)|^{2} \ra$,
$\la \chi_{1}^{x}\ra=\la |s(2\pi/L_{x},0,0)|^{2} \ra$,
$\la \chi_{1}^{z}\ra=\la |s(0,0,2\pi/L_{z})|^{2} \ra$
thermal average  of Fourier components for particular
realization of impurities.
Then we can compute the corresponding correlation lengths
averaged over impurity realizations
\begin{eqnarray}
\xi_{\perp}{=}\frac{1}{2 \sin( \pi/L_{x})}
\overline{\sqrt{\frac{\la \chi_{0}\ra}{\la \chi_{1}^{x}\ra}-1}},\;\;
\xi_{||}{=}\frac{1}{2 \sin( \pi/L_{z})}
\overline{\sqrt{\frac{\la \chi_{0}\ra}{\la \chi_{1}^{z}\ra}-1}}.\nonumber\\
\end{eqnarray}

The simulation is performed with hybrid MC method.~\cite{LB}
Each of MC step consists of one flip of Wolff cluster
followed by $V/4$ attempts to flip
spins in accordance with Metropolis rule.

In our study we deal with a concentration of impurities $p=0.2$. Such concentration is very often used in MC simulations of Ising model with disorder,\cite{Ballesteros99,Prudnikov_MC,Ivaneyko08,Ballesteros98} since   in this case concentration of magnetic sites  $1-p=0.8$ is far from the 
percolation threshold and from the pure system. 
Additional empirical reason to take this value of $p$ is the  following: for 3d Ising model with uncorrelated impurities, correction-to-scaling terms were found to be minimal at concentration of magnetic sites $0.8$.\cite{Ballesteros98}

\section{Results and analysis of simulations}\label{V}
In this section, applying the above described formalism, we give three estimates of the anisotropy exponent obtained from anisotropic FSS predictions for different quantities. From the FSS of susceptibility we get also an estimate for $\gamma/\nu_{\perp}$.

\subsection{Computation of the correlation length}
To use relation (\ref{relat}) we need to perform calculations at the 
critical temperature of an infinite system.  
Let us assume that the value of the critical temperature of an 
infinite system is very close to the value of the 
pseudo-critical temperature of the system with maximal size.  In this case that is the temperature of the susceptibility maximum 
$\chi_{\mathrm{max}}$.
We perform the computation for a cubical system of size $128^{3}$
with concentration  of impurities $p=0.2$.   Disorder average may be performed in either of two  possible protocols: In the first procedure, at each disorder realization, the temperature dependence  of the susceptibility is obtained. Then all these curves are averaged to get a single curve of $\chi$ that depends on $\beta$. Then the value $\beta$ at which $\chi(\beta)$ has a maximum  is associated with the critical temperature  (we refer this way as {\em method A}). 
The second alternative is to average temperatures at which
$\chi_{\mathrm{max}}$ is achieved for each disorder realization
(we refer this way as {\em method B}). 

\begin{figure}
\includegraphics[width=0.44\textwidth]{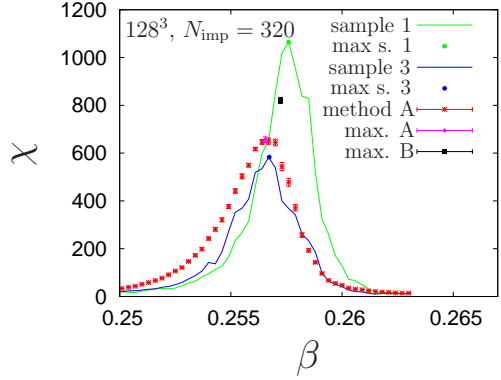}
\caption{
(Color online) Maxima of magnetic succeptibility $\chi$
 $\beta_{\mathrm{ max}}^{A}=0.2565$ for {\em method A} and
 $\beta_{\mathrm{max}}^{B}=0.2572$ for {\em method B} for a system $128^{3}$
 and $p=0.2$.
}
\label{fig:ex_xi}
\end{figure}

The results obtained with both methods are given in
 Fig.~\ref{fig:ex_xi}. The two curves displayed by continuous lines in the figure show $\chi(\beta)$ behaviour for two separate samples, whereas the dotted curve shows the $\chi(\beta)$ averaged over 320 samples.
Using the {\em method A} we first average the susceptibility
and then find the point $\beta_{\mathrm{max}}^{A}=0.2565$
-- magenta diamond   in Fig.~\ref{fig:ex_xi}. 
Using the {\em method} B we first find a maximum of susceptibility
for each disorder configurations, and then average location of these maxima
over samples
 $\beta_{\mathrm{max}}^{B}=0.2572$
-- black square in Fig.~\ref{fig:ex_xi}. 

\begin{figure}
\includegraphics[width=0.49\textwidth]{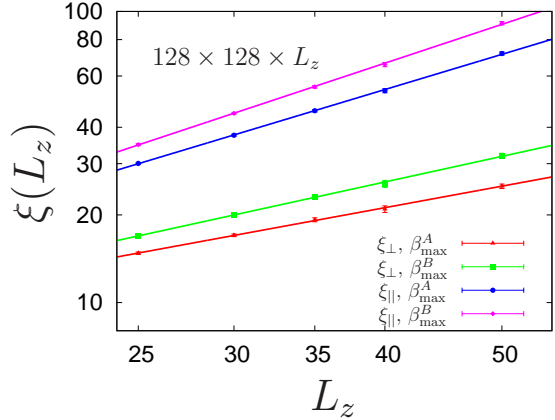}
\caption{
(Color online)
Results of the fit of $\xi$ as a power law dependence  of $L_{z}$ for a system of size
 $128 \times 128 \times L_{z}$ and $p=0.2$
at points  $\beta_{\mathrm{ max}}^{A}=0.2565$ for {\em method A} and
 $\beta_{\mathrm{max}}^{B}=0.2572$ for {\em method B}.
}
\label{fig:xim3}
\end{figure}

With the value of the critical temperature at hand  we can perform a study through relation (\ref{relat}) to extract the anisotropy exponent $\theta$.
To this end, we analyze the correlation length $\xi_{\perp}$ of the system of size $128 \times 128 \times L_{z}$ varying $L_z$ for both  values of 
$\beta_{\mathrm{max}}$.
Results of simulations for $L_z = 25 - 50$ are given in Fig.~\ref{fig:xim3},
where we plot $\xi$ for $\perp$ and $||$ directions as function of $L_z$.

Then we perform the fit of data for $\xi_{\perp}$ in accordance with formulae:
\begin{equation}
\label{eq:fit_xi}
\xi_{\perp}(L_{z})= a L_{z}^{b},
\end{equation}
using fitting parameters $a$, $b$. For {\em method  A} we get
$a=1.27(4)$,  $b=0.76(1)$, while for {\em method B} we estimate $a=0.91(3)$,  $b=0.90(1)$.
Comparing (\ref{eq:fit_xi})  with (\ref{relat}) we get $\theta=1/b$ with the following estimates: $\theta\approx 1.31$ ({\em method A}) and $\theta\approx 1.11$ ({\em method B}).
Taking these values as an accuracy interval of $\theta$ determining, $1.1\lesssim\theta\lesssim 1.3$ we see that they are in reasonable agreement with existing analytic RG estimates (see Section \ref{III}). 

\subsection{Computation of the Binder cumulant}
\label{subsec:VB}
Another way to identify the right value of
the exponent $\theta$ is the following.
We expect that all curves for the temperature dependence of the Binder cumulant $U_{4}$  for systems of different sizes but fixed generalized aspect ratio with proper $\theta$ will intersect  at the critical point.
Therefore we can consider the system of size $L_{x}\times L_{y} \times L_{z}$
with $L_x=L_y$ and keeping  $L_{z}=L_{x}^{\theta}$ for various values of $\theta$.
For this system we compute the cumulant $U_{4}(\beta,L_{x})$
as a function of the inverse temperature $\beta$
for various values of  $L_{x}$.
At the proper value $\theta^{*}$ we expect the intersection of graphs.
We perform simulations for $L_{x}=20,40,60,80,100$,
the averaging is performed over $N_{\rm imp}=128$ realization of impurities.

\begin{figure*}[htbp]
\includegraphics[width=0.44\textwidth]{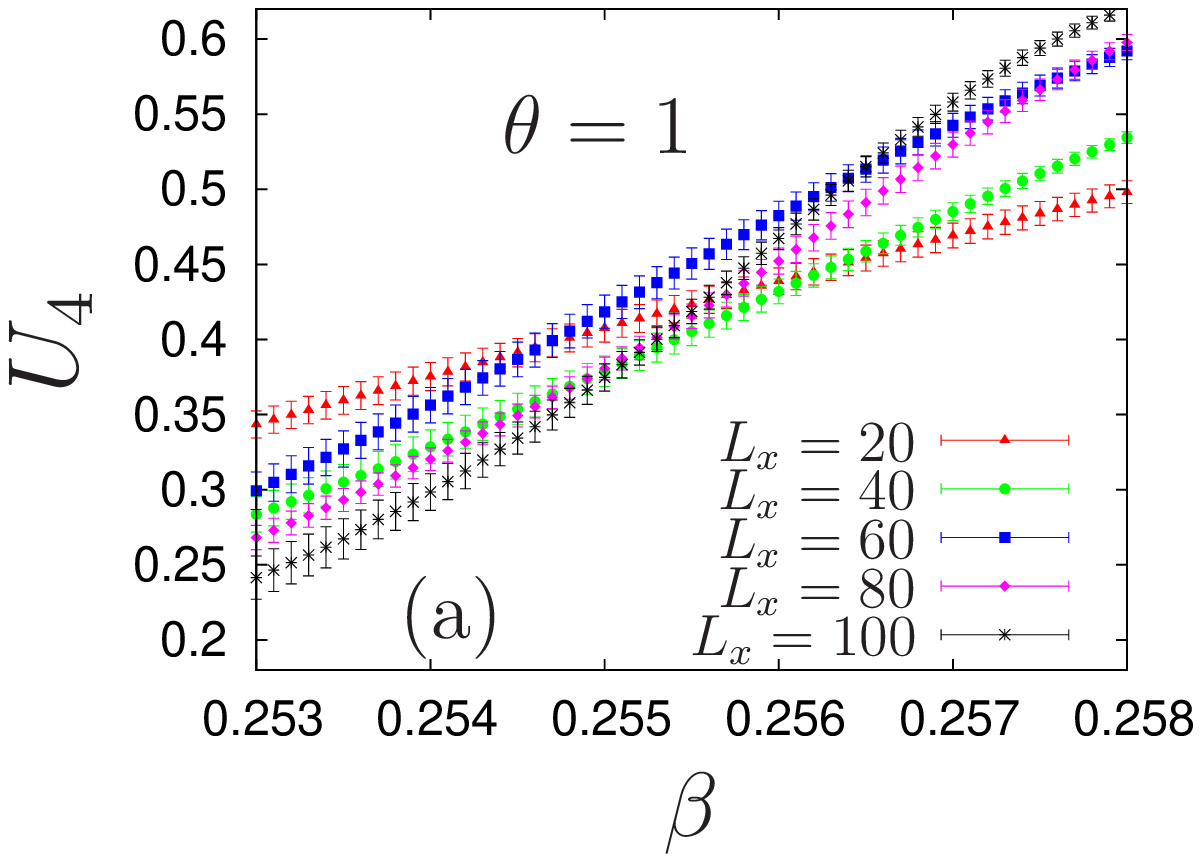}
\includegraphics[width=0.44\textwidth]{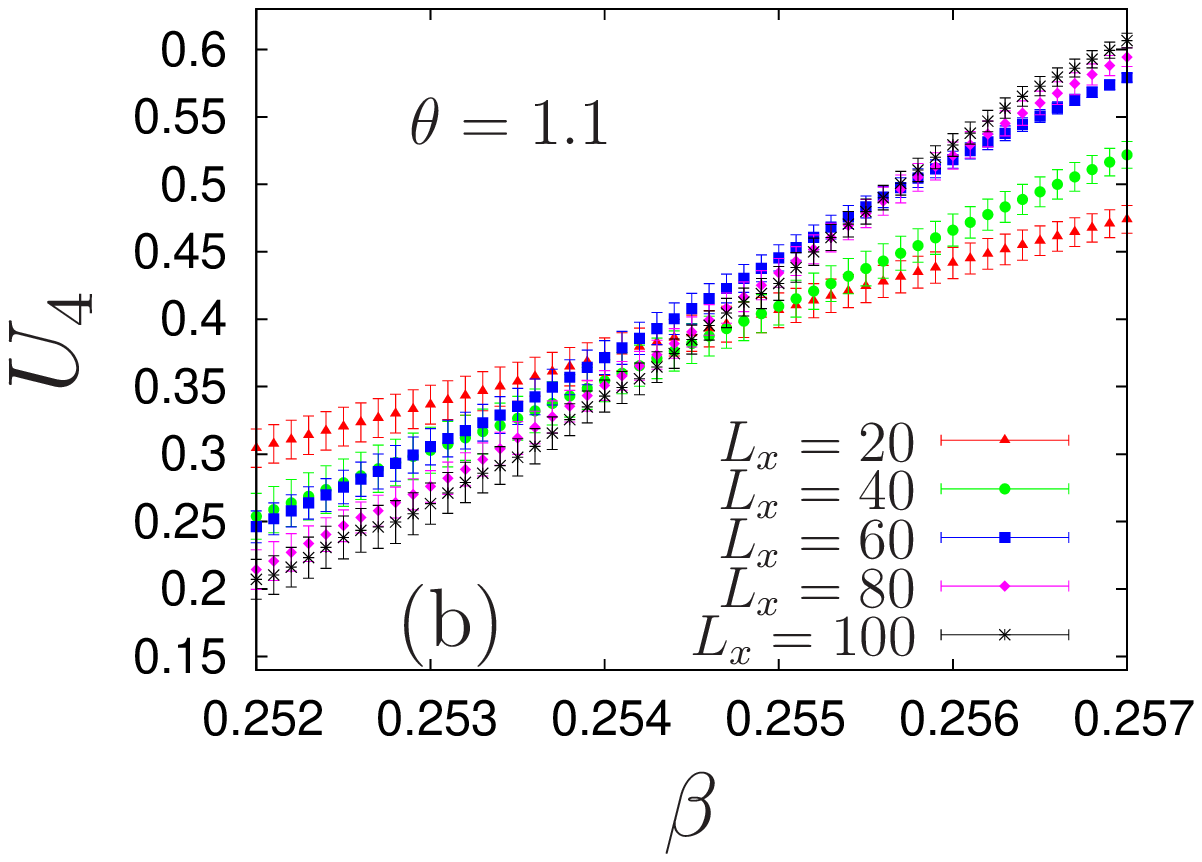}
\includegraphics[width=0.44\textwidth]{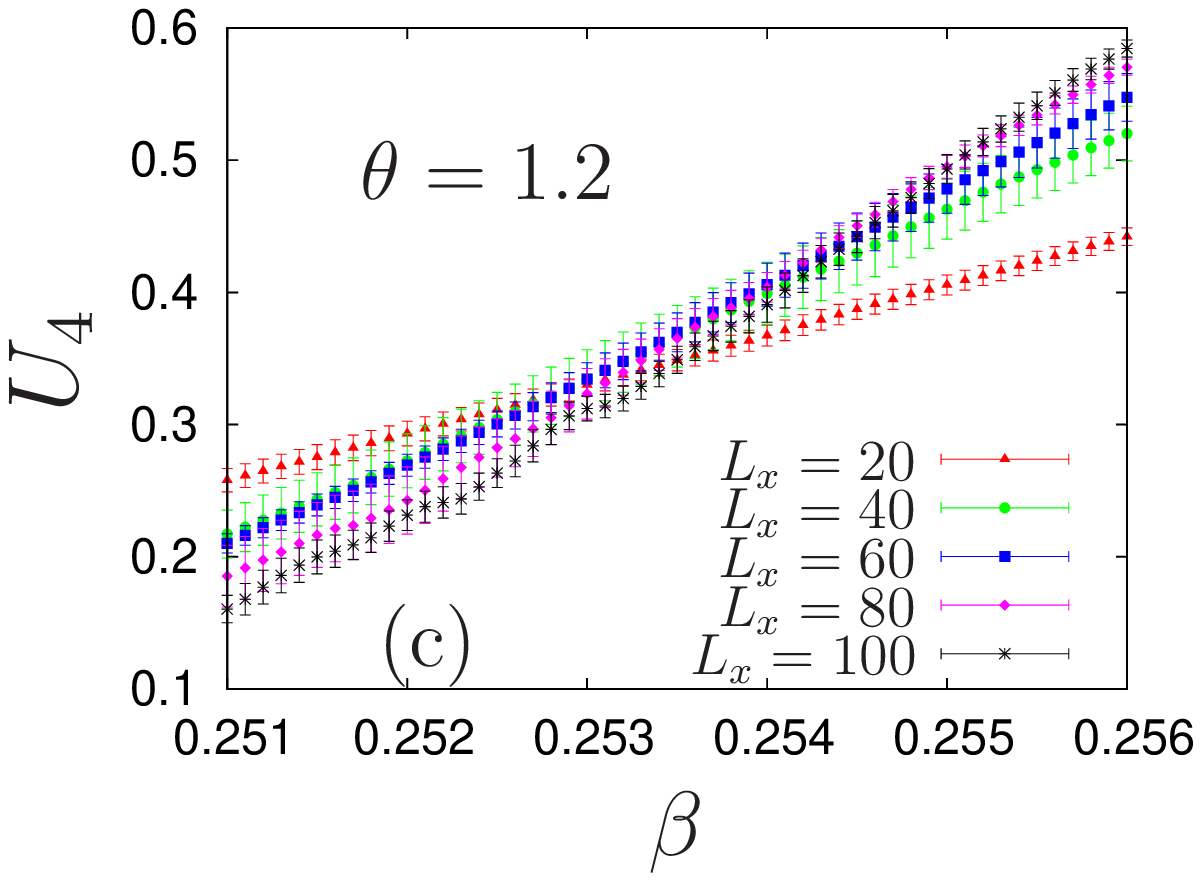}
\includegraphics[width=0.44\textwidth]{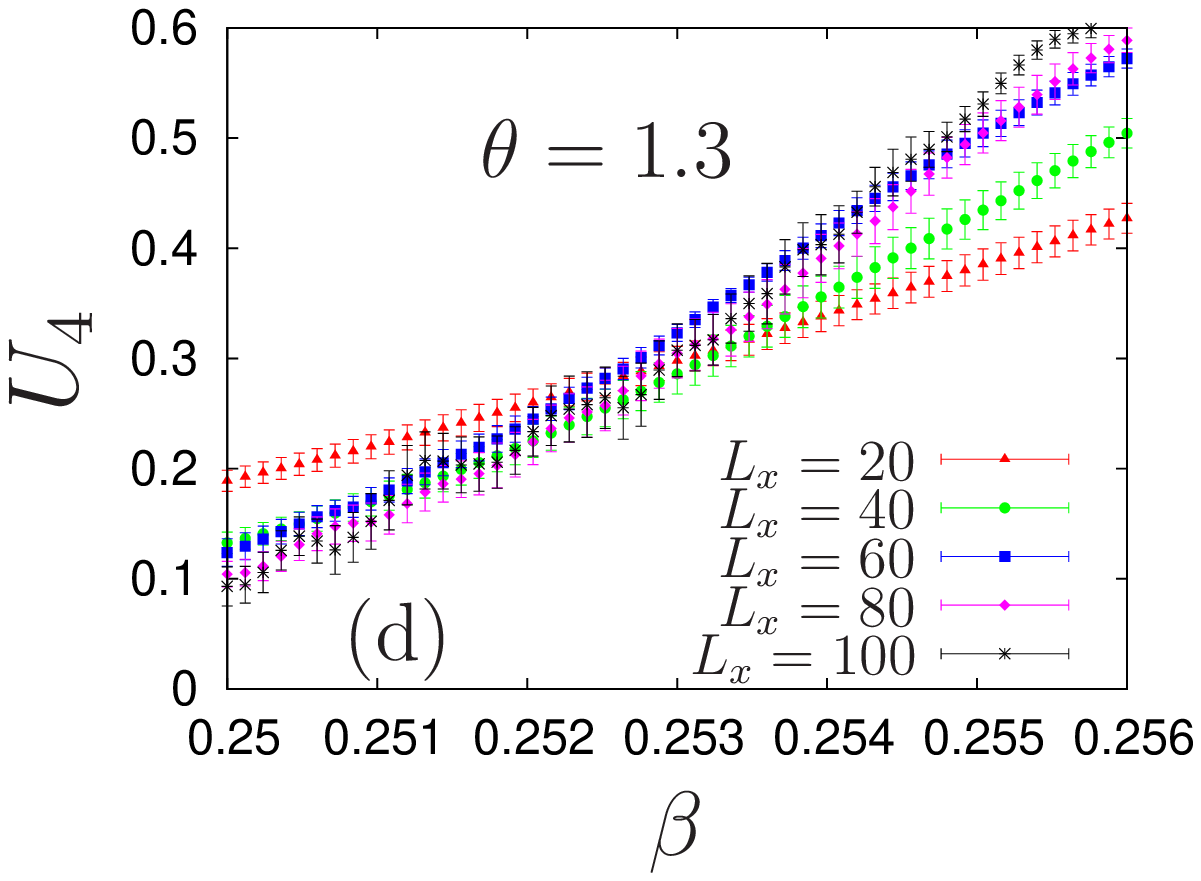}
\caption{
(Color online)
Magnetization cumulant $U_{4}$ as a function of the inverse temperature $\beta$
for $L_{z}=L_{x}^{\theta}$:
(a) $\theta=1.0$;
(b) $\theta=1.1$;
(c) $\theta=1.2$;
(d) $\theta=1.3$.
}
\label{fig:u4}
\end{figure*}

\begin{figure*}
\includegraphics[width=0.49\textwidth]{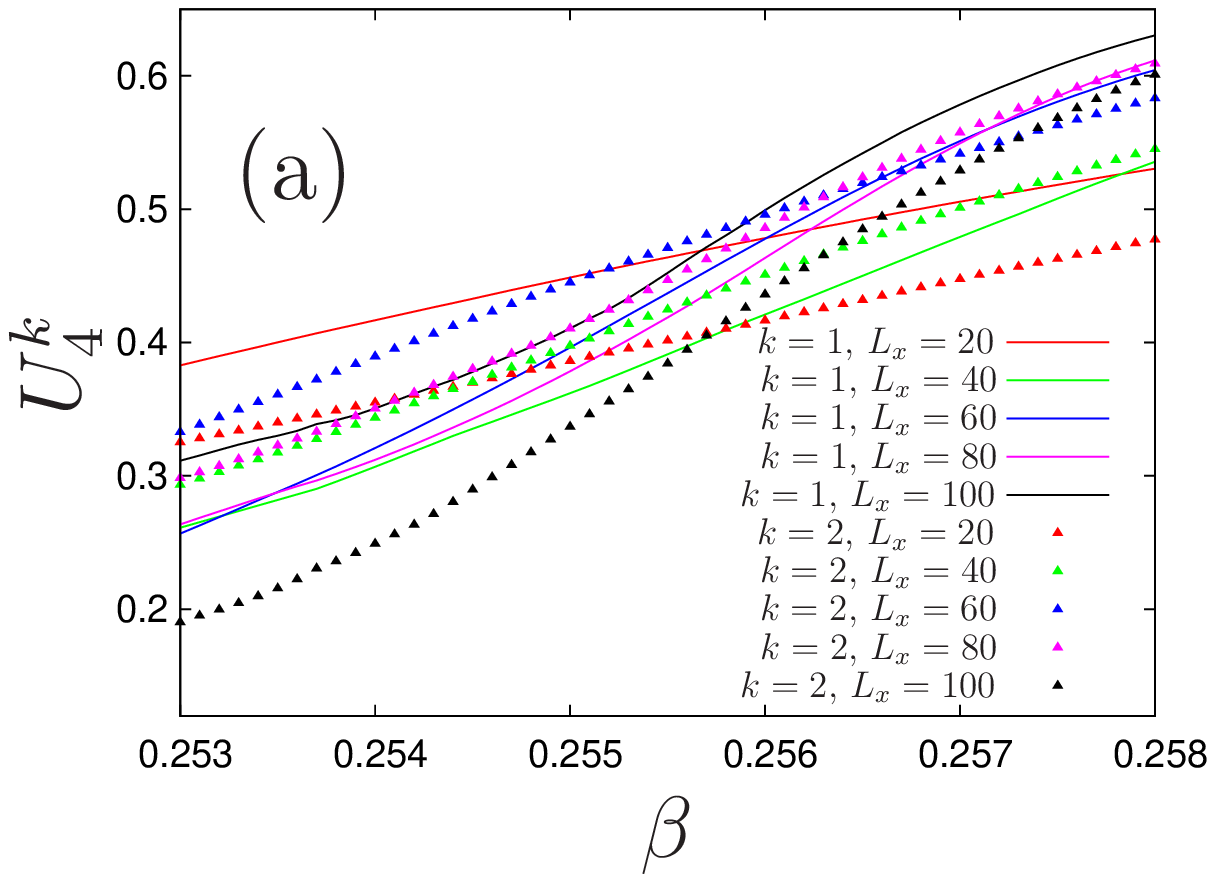}
\includegraphics[width=0.49\textwidth]{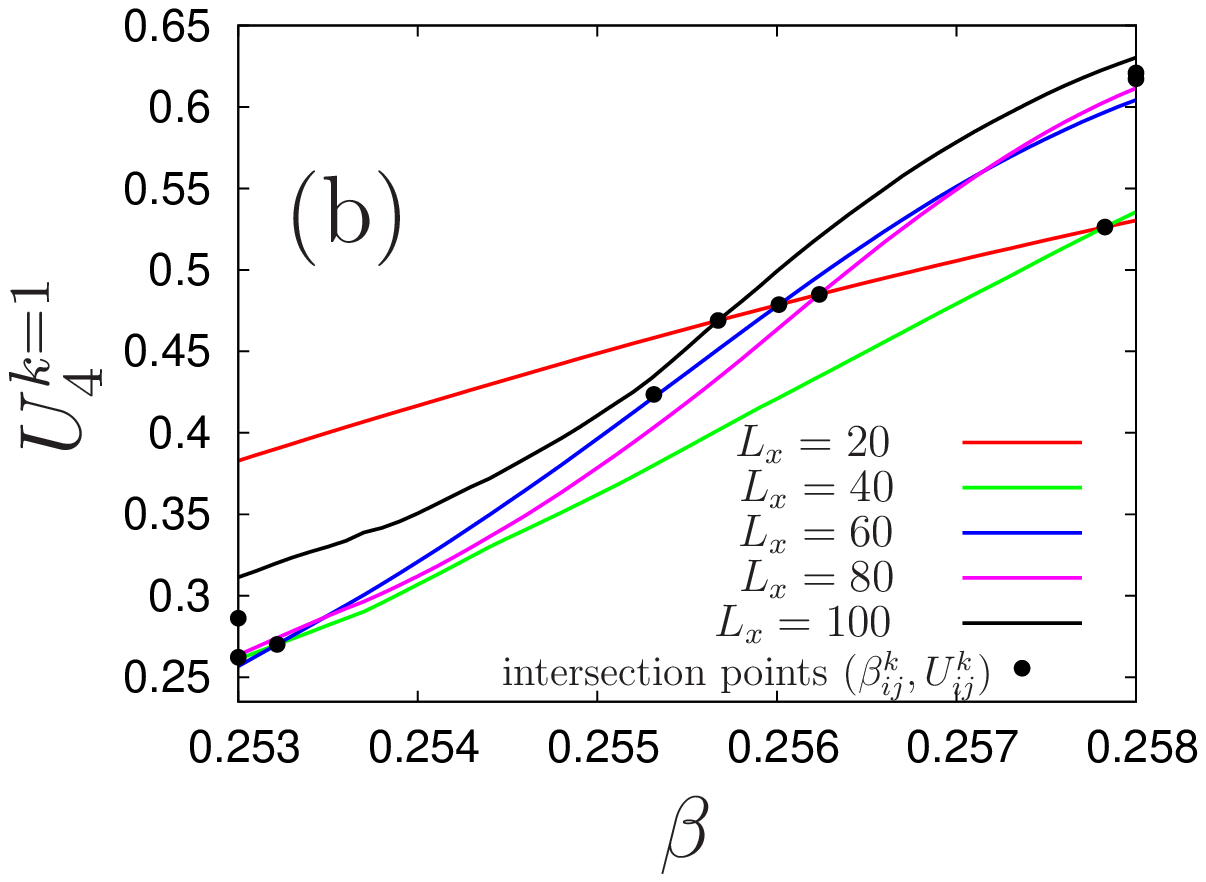}
\caption{
(Color online)
The cumulant $U_{4}^{k}$ for $k$-th series as a function of
$\beta$ for $\theta=1$ and various values of $L_{x}=20,\ 40,\ 60,\ 80,\ 100$:
(a) Comparison of results for $k=1$ (lines) and $k=2$
(symbols). Note, that some pair of graphs
(for example, for $L_{x}=40$ and $L_{x}=80,\ 100$ for $k=1$)
 have no intersection within the considered range.
(b) Results for $k=1$ and intersection points $(\beta_{ij}^{k},U_{ij}^{k})$.
}
\label{fig:testu}
\end{figure*}

In Fig.~\ref{fig:u4}(a)-(d)
we plot the cumulant for values $\theta=1,\ 1.1,\ 1.2,\ 1.3$,
respectively. Each curve corresponds to $U_4(\beta)$ averaged over 128 disorder realizations.
As one can see, the presence of disorder smears the  single crossing point into a region of temperatures where all curves cross.
The narrowest region is expected for the value of $\theta$ which is sufficiently close to the real one.
To  analyze  this situation for different values of $\theta$, we use
the following procedure.   We split the total set of
$N_{\mathrm{imp}}=128$ different impurity realizations onto
eight series, and compute the average value (averaged over
$128/8=16$ realizations) of the
invariant $U_{4}^{k}$ for each set $k=1,2,\dots,8$.
Later on we use the averaging over $U_{4}^{k}$ for
evaluation of numerical inaccuracy.
In Fig.~\ref{fig:testu}(a)
we plot for comparison results for $U_{4}^{k}$ as a function of $\beta$
for two different series $k=1$ (lines) and $k=2$ (triangles)
for $\theta=1$ and various sizes $L_{x}=20,\ 40,\ 60,\ 80,\ 100$.
 In this figure we observe an important difference between
graphs for two series  due to impurity induced fluctuations.
 Our aim is to study the scattering of intersection
 points. Ideally, graphs for five system sizes should intersect in ten points.
 In Fig.~\ref{fig:testu}(b) we plot data for $U_{4}^{k=1}$
 and indicate the intersection points $(\beta_{ij}^{k},U_{ij}^{k})$
 by black circles. Here a pair $ij$ of indexes labels
 two system sizes $i=1,\ 2,\ 3,\ 4,\ 5$ for $L_{x}=20,\ 40,\ 60,\ 80,\ 100$.
  The graphs for some  system
 sizes  have intersections outside of
 considered interval for $\beta$, {\it e.g.}, for $L_{x}=40$
 and for $L_{x}=80,\ 100$. In this case we select as
 ``intersection point'' the point between left side or right side
 ends of these two lines selecting the side with the smaller distance
 between the endpoints.

Another possible situation with multiple intersection points
may happen due to big scattering of the points in graphs caused by
numerical inaccuracy. In this case we perform the averaging
over all intersection points and proceed with this averaged point.
Then we compute the average values of the inverse temperature
$$
\beta_{\mathrm{av}}^{k}=\frac{1}{10}
\sum \limits_{i}^{5} \sum \limits_{j < i}^{5} \beta_{ij}^{k}
$$
and of the cumulant
$$
U_{\mathrm{av}}^{k}=\frac{1}{10}
\sum \limits_{i}^{5} \sum \limits_{j < i}^{5} U_{ij}^{k}
$$
over ten intersection points for five system sizes for each set $k$.
We observe (as expected), that average of crossing points
do not coincide with crossing of average lines.

\begin{figure*}
\includegraphics[width=0.49\textwidth]{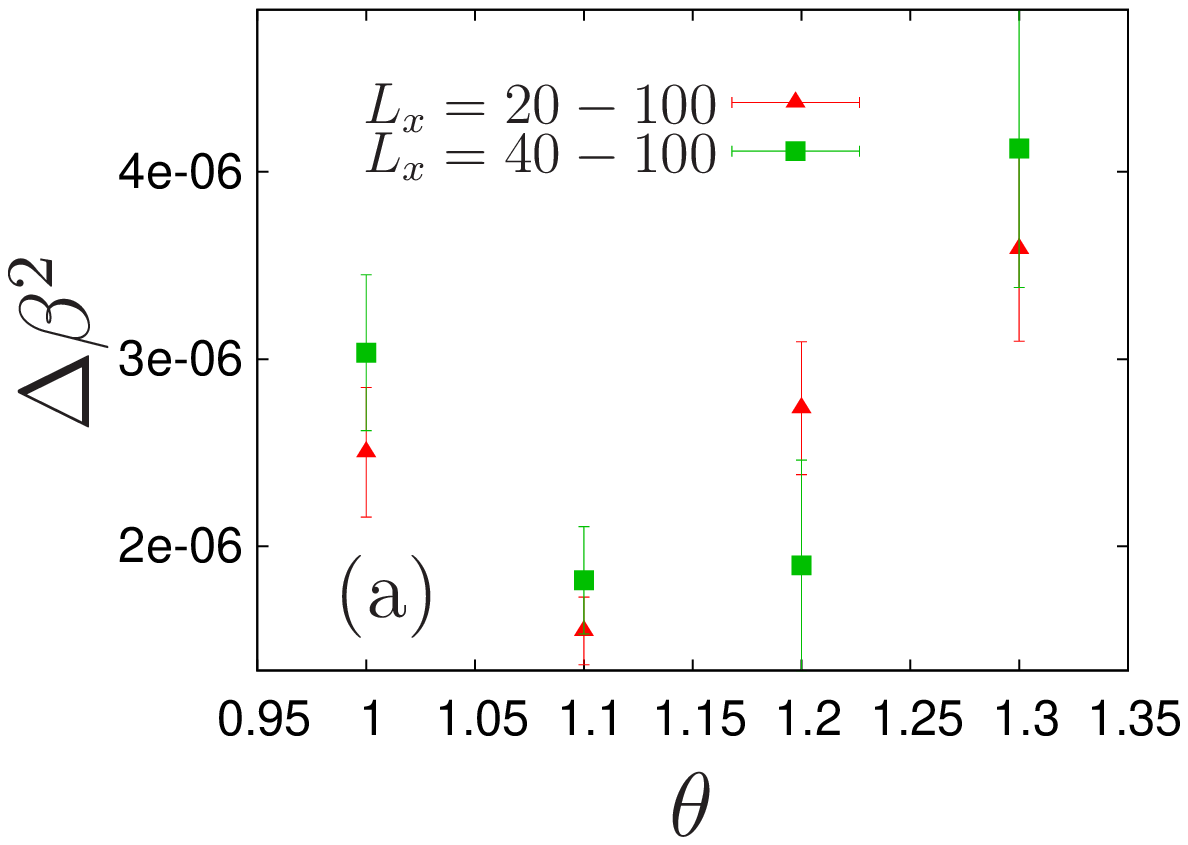}
\includegraphics[width=0.49\textwidth]{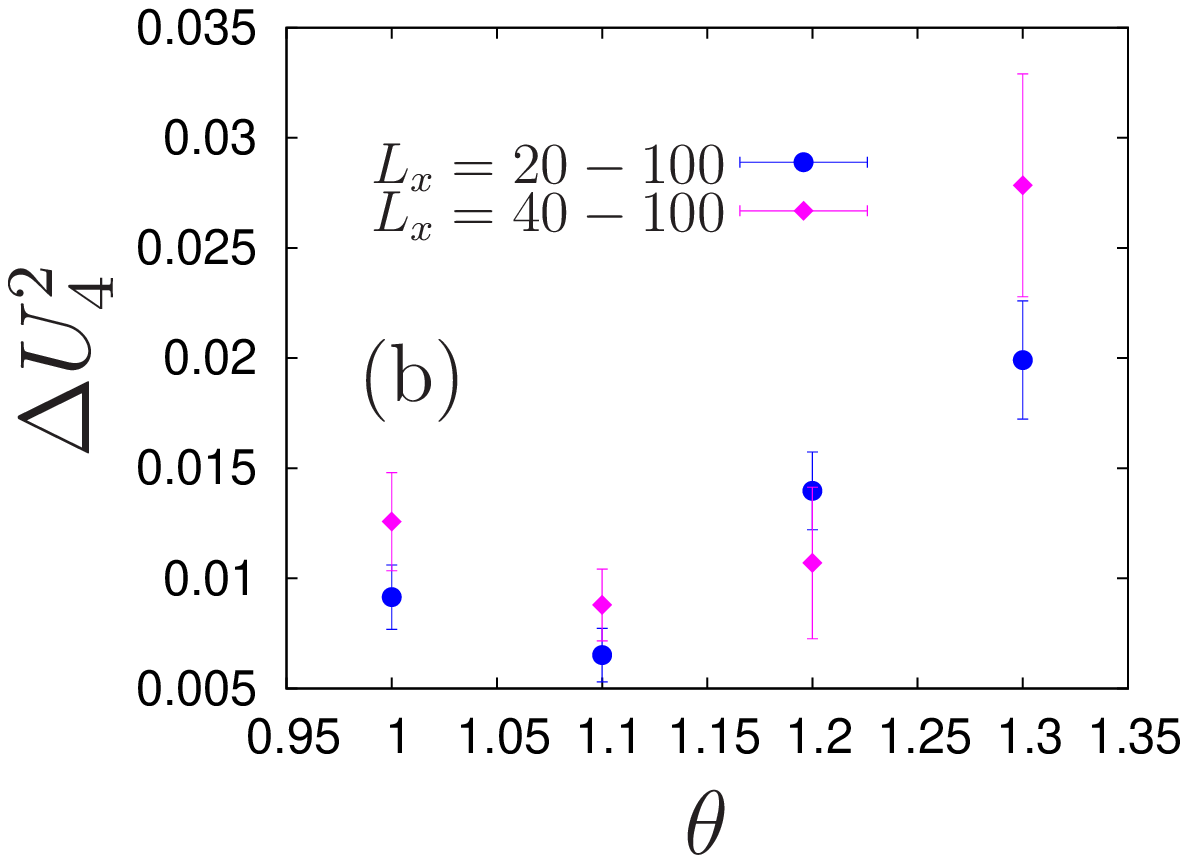}
\caption{
(Color online)
(a) The average of the square deviation of the inverse temperature
 $\Delta \beta^{2}$ of crossing points as a function of $\theta$;
(b) The average of the square deviation of the cumulant
 $\Delta U_{4}^{2}$ of crossing points as a function of $\theta$.
 }
\label{fig:dbetau}
\end{figure*}

Now we can compute the average values
of the square of deviations  from the mean values for the inverse temperature
$$
\Delta \beta^{2}=\frac{1}{8 \times 10}
\sum \limits_{k=1}^{8}\sum \limits_{i}^{5} \sum \limits_{j < i}^{5} (\beta_{ij}^{k}-\beta_{\mathrm{av}}^{k})^{2}
$$
and for the cumulant
$$
\Delta U_{4}^{2}=\frac{1}{8 \times 10}
\sum \limits_{k=1}^{8}\sum \limits_{i}^{5}
\sum \limits_{j < i}^{5} (U_{ij}^{k}-U_{\mathrm{av}}^{k})^{2}
$$
and evaluate numerical inaccuracy.
In Fig.~\ref{fig:dbetau}(a) we plot $\Delta \beta^{2}$
as a function of $\theta$.
The amplitude of the square of the inverse temperature deviation
is about $10^{-6}$ which is consistent with the average distance
between points $\Delta \beta \sim 10^{-3}$.
We do not observe pronounced minima in the interval $1.0 \le \theta \le 1.3$,
and the numerical inaccuracy is comparable with the scattering of the points.
In Fig.~\ref{fig:dbetau}(b) we plot data for the average square
of the deviation for the cumulant $\Delta U_{4}^{2}$.
The point distribution  is very similar to the ones obtained for the inverse temperature.
The point for $\theta=1.1$ is below its neighbors, but
the distance between these points is of the order of the
numerical inaccuracy.
Therefore the above described procedure corroborates $\theta=1.1$ as the optimal one.

\subsection{Computations of the susceptibility}

\begin{figure*}
\includegraphics[width=0.49\textwidth]{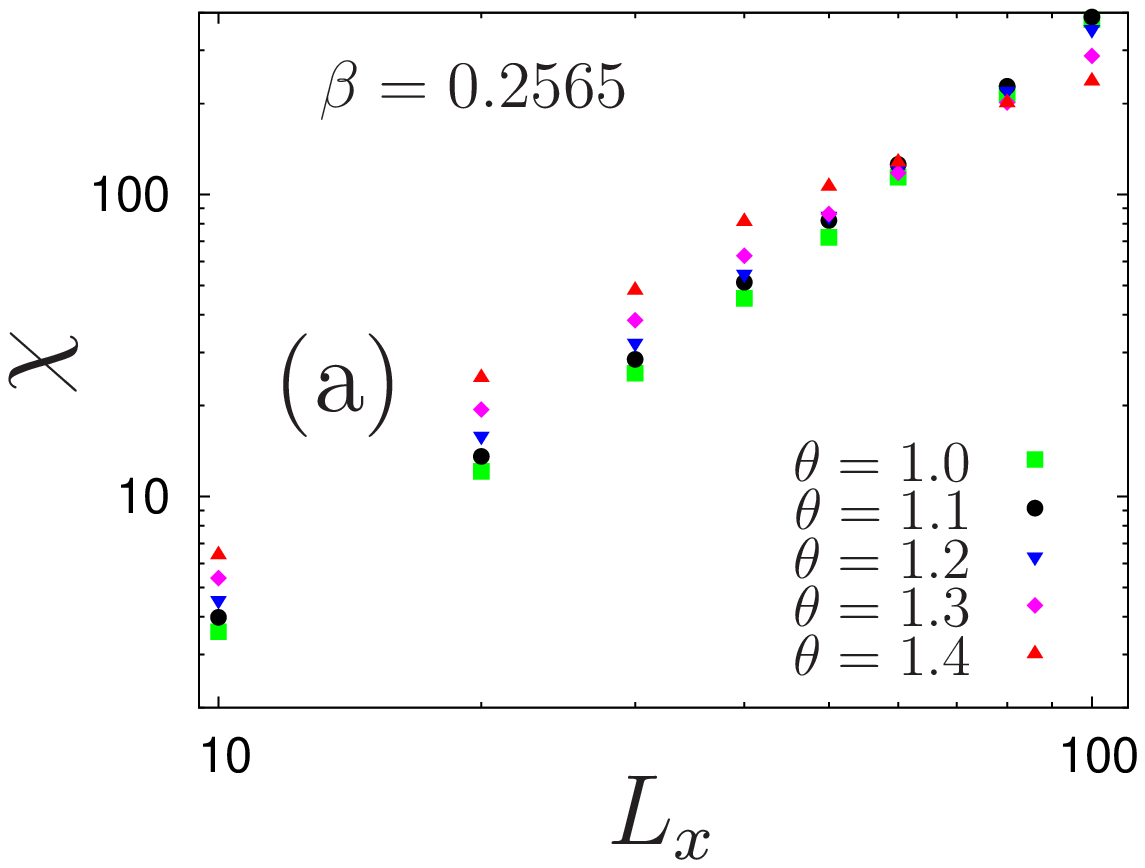}
\includegraphics[width=0.49\textwidth]{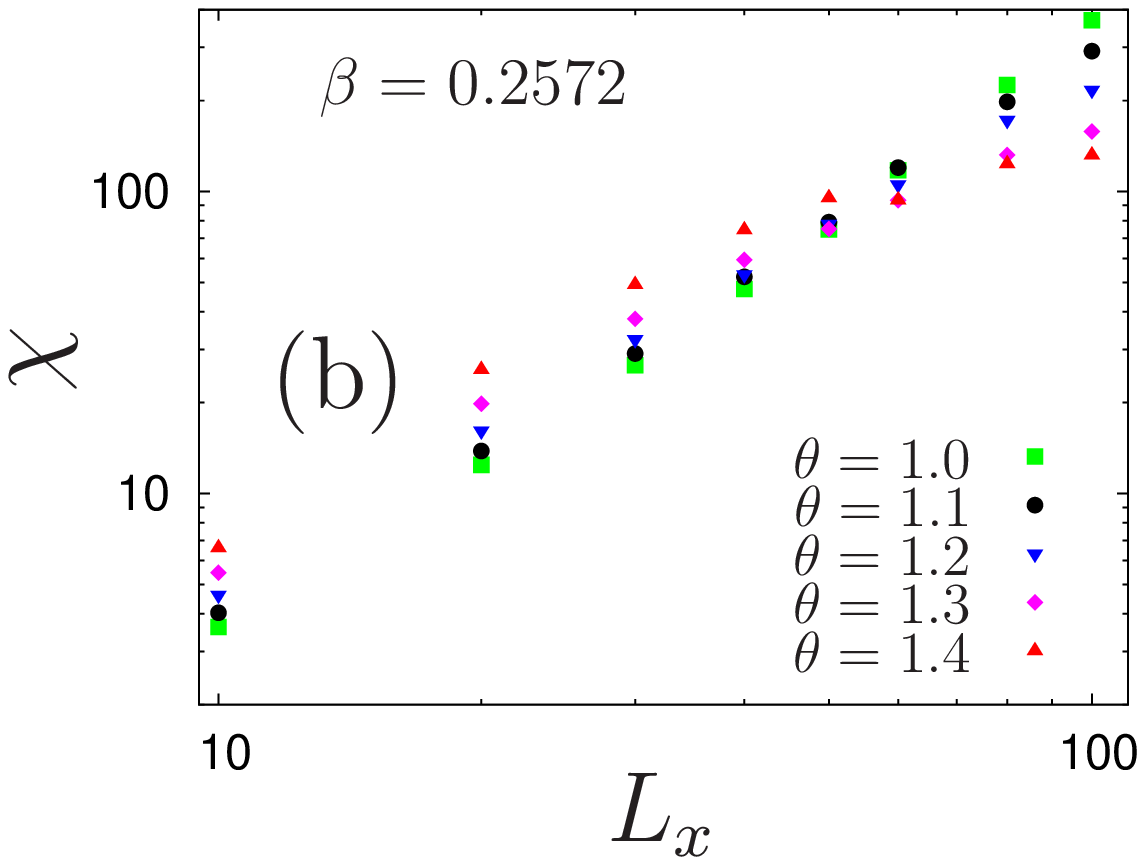}
\caption{
(Color online)
Magnetic susceptibility $\chi$ as a function of the system size
$L_{x}$ for various values of $\theta=1,\ 1.1,\ 1.2,\ 1.3,\ 1.4$
at the point:
(a) $\beta_{\mathrm{max}}^{A}=0.2565$,
(b) $\beta_{\mathrm{max}}^{B}=0.2572$.
 }
\label{fig:chi_com_l}
\end{figure*}

In this subsection we describe the results of the computation of the
magnetic susceptibility at the  critical point $\beta_{c}$
 with various values of the exponent $\theta$.
Before we obtained two estimates for the critical
$\beta_{c}$:
 $\beta_{\mathrm{max}}^{A}=0.2565$ (for {\em method A})
 and $\beta_{\mathrm{max}}^{B}=0.2572$ (for {\em method B}).
In Figs.~\ref{fig:chi_com_l}(a)~and~(b)
we plot the magnetic susceptibility $\chi$ as a function of the system size
$L_{x}$ in the log-log scale
for $\beta_{\mathrm{max}}^{A}$
and $\beta_{\mathrm{max}}^{B}$, respectively.
We expect the power law behavior $\chi(L_{x}) \propto L_{x}^{\gamma/\nu_{\perp}}$
for the more satisfying  value of the exponent $\theta$.
In Fig.~\ref{fig:chi_com_l}(a) the deviation of the 
data from a straight line is less than in Fig.~\ref{fig:chi_com_l}(b)
therefore we select the estimate $\beta_{c}=\beta_{\mathrm{max}}^{A}=0.2565$
for future analysis. This is consistent with the analysis of Binder cumulants
where it appears that $\beta^{B}_{\mathrm{max}}$ 
seems to be out of the crossing window of $U_{4}^{k}$.

We perform a fit of the magnetic susceptibility
 by a linear
function of $L_{x}$ in the log-log scale: $\chi = aL_x^b$
 with $a$ and $b$ being the parameters of the fit.
Then we  estimate the deviation of the values of susceptibility measured numerically from those obtained via the fitting function with the help of a chi-square   
$\hat\chi^{2}$ defined as
\begin{equation}
\label{eq:chi2}
\hat\chi^2=\sum_{i=1}^N
\frac{(\chi_i-f(L_i))^2}{\sigma_i^2}
\end{equation}
where $N=8$ is the number of  values  
$\chi_i$ calculated for $i=1,2,\dots, N$ values of $L_i=10,20,30,40,50,60,80,100$
the system size $L_x$, 
 $f(x)$ is the fitting function, 
 $\sigma_{i}^2$ is the appropriate variance defined by the error bars.
 
The same number of MC steps is used
for a given value of the exponent $\theta$.
Therefore, the variance $\sigma_{i}$
is minimal for small values  of the system size $L_{x}$
and increases for larger values of $L_{x}$.
  The total number of MC steps decreases with increasing of $\theta$
  (from $5 \times 10^{6}$ for $\theta=1$ to $5 \times 10^{5}$
  for $\theta=1.45$).

\begin{figure}
\includegraphics[width=0.49\textwidth]{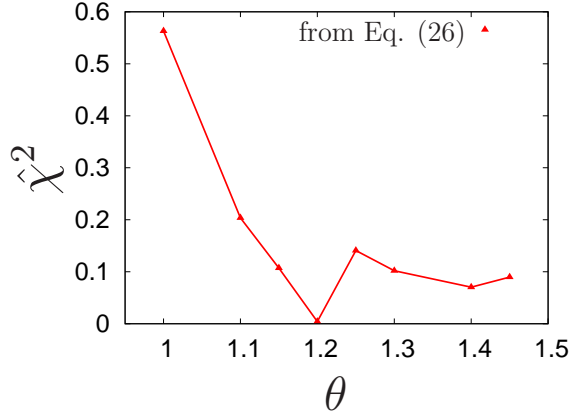}
\caption{
(Color online)
The $\hat\chi^{2}$ parameter  of the deviation from 
 the linear fit (Eq.~(\ref{eq:chi2}))
of the magnetic susceptibility
as a function of $\theta$.
 }
\label{fig:chi2_err}
\end{figure}

\begin{figure*}
\includegraphics[width=0.49\textwidth]{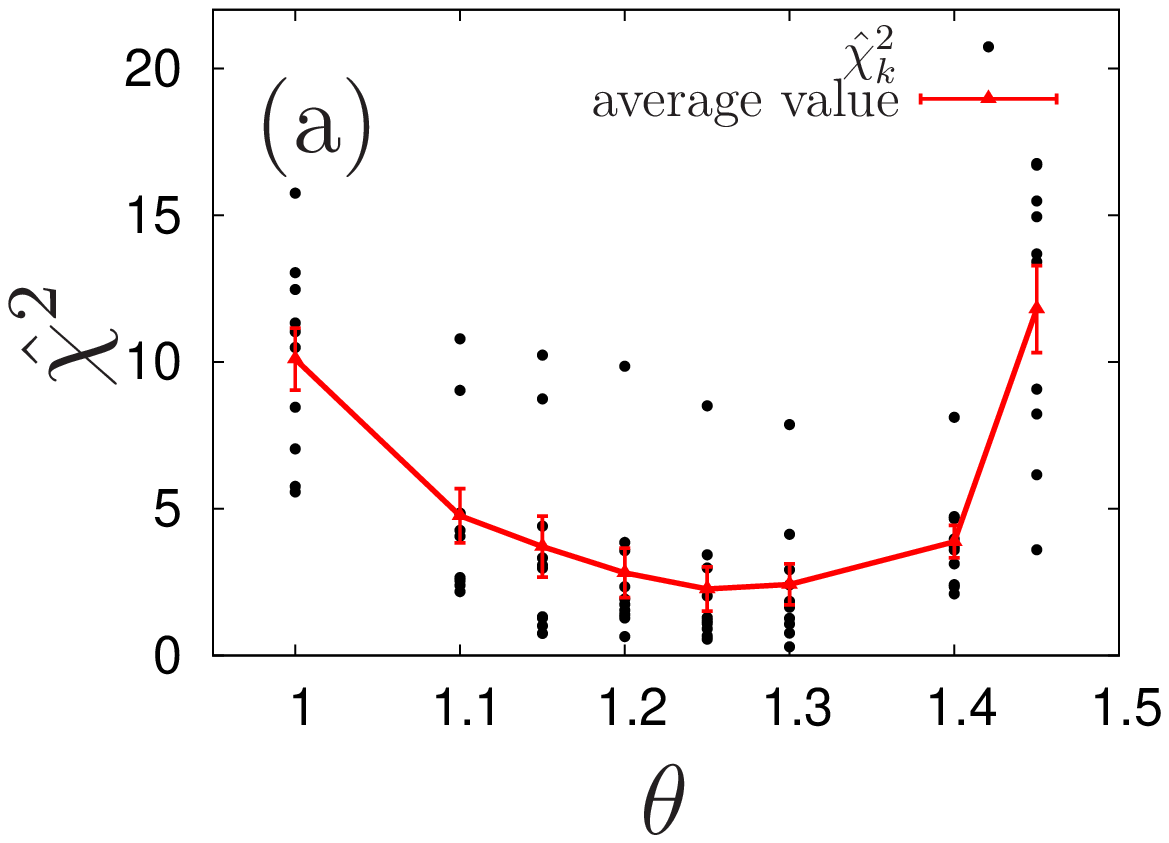}
\includegraphics[width=0.49\textwidth]{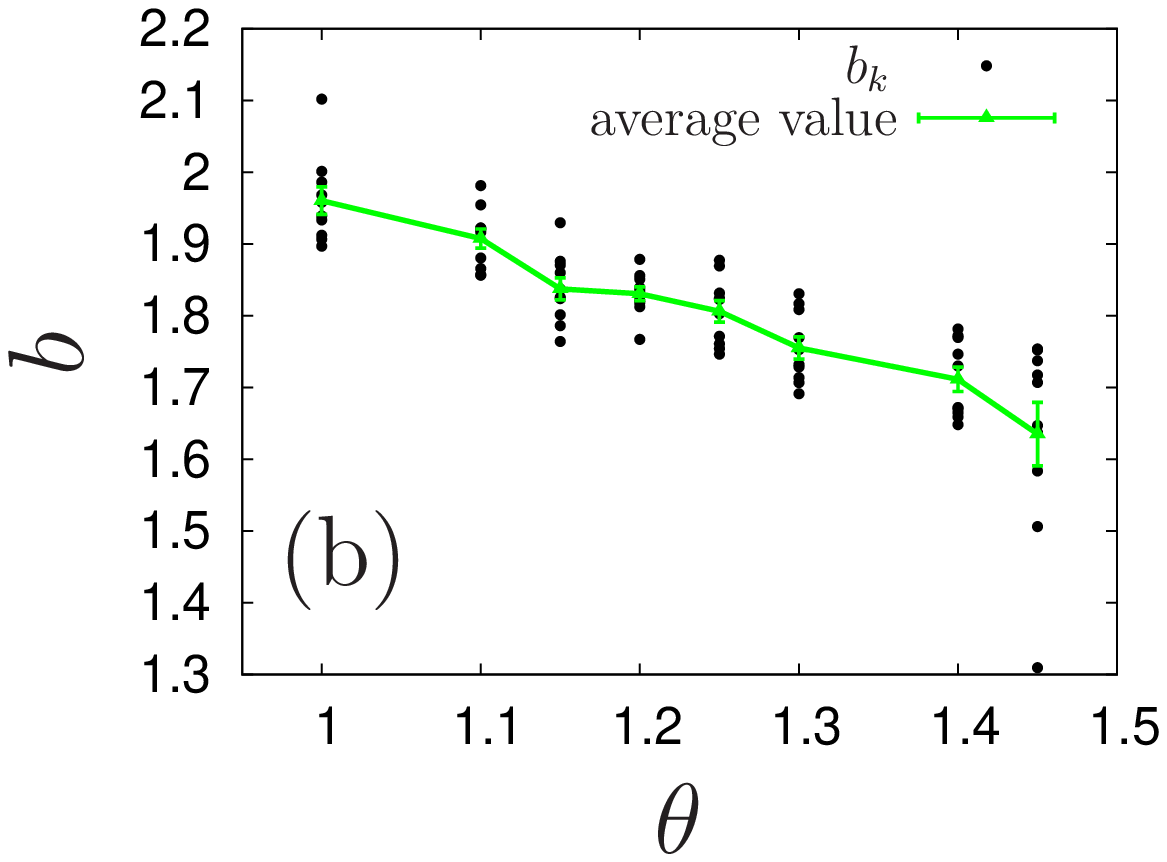}
\caption{
(Color online)
(a) The $\hat\chi^{2}$ parameter 
 of the linear fit with variance
($\hat\chi_{k}^{2}$ for every series black circles, average result
red triangles)
as a function of $\theta$;
(b) The result of the power of the fit $b$
($b_{k}$ for every series black circles, average result
green triangles) as a function of $\theta$.
 }
\label{fig:res_err}
\end{figure*}

The parameter $\hat\chi^{2}$ of the fitting procedure
characterizes the ``quality'' of the data with respect to the
 proposed functional dependence. In our case this parameter
describes the deviation of points from the straight line in the log-log representation.
We plot $\hat\chi^{2}$ as a function of $\theta$
in Fig.~\ref{fig:chi2_err}.
The general tendency is the decrease of
$\hat\chi^{2}$ with an increase of $\theta$, because due to the smaller
number of MC steps the variance is larger in this case.
We observe a minimum in the region $\theta \approx  1.2-1.3$.
Unfortunately, for this procedure we cannot evaluate the
variance (inaccuracy) $\delta \hat\chi^{2}$. Therefore we repeat the procedure in more
regular way.
We split 320 impurity realizations (for each value of $L_{x}$
for a fixed $\theta$) into 10 series and perform the fit by the formula
$\ln(\chi) \simeq \ln(a_{k})+ b_{k} \ln(L_{x})$ for every series obtaining some value
$\hat\chi^2_{k}$. Then we average these values, compute numerical inaccuracy
and plot the results in Fig.~\ref{fig:res_err}(a)
(black circles for $\hat\chi_{k}^{2}$ and red triangles for the average value).
We can see, that $\hat\chi^{2}$ reaches the minimum
for $\theta^{*} \simeq 1.25$. In Fig.~\ref{fig:res_err}(b)
we plot the resulting power $b$ as a function of $\theta$.
Value of $b$ at $\theta^*$  gives an estimate for 
$\gamma/\nu_{\perp}\simeq 1.85\pm 0.05$, using the window $1.1\le\theta\le 1.3$.
We present here also estimate 
$\gamma/\nu_{\perp}=1.90\pm 0.08$  
obtained from data of Fig.~\ref{fig:res_err} 
for the value $\theta=1.1$ recognized as an optimal value in 
subsection~\ref{subsec:VB} at cumulant analysis.

\section{Conclusion}\label{VI}
In this paper we have studied  by MC simulations the scaling  
behaviour  of thermodynamical quantities in the vicinity of critical 
point for 3d Ising system with randomly distributed parallel  linear
extended defects, modeled as non-magnetic impurities collected into 
lines along spatial direction $z$. We considered combined algorithm 
using Wolff and Metropolis methods. Our results are consistently 
interpreted using the theory of anisotropic finite size scaling. 

We have estimated the   value of the anisotropy exponent $\theta$ 
using three different methods, namely, from dependence of correlation 
length $\xi_{\perp}$  on the linear size of the system near the critical point, 
from a temperature dependence of the fourth-order Binder's cumulant, 
from finite-size scaling of the susceptibility. The values estimated 
are in the range $1.1\le\theta\le 1.3$ and corroborate RG predictions for the  
model under consideration.

We have  also measured the value  of $\gamma/\nu_{\perp}$  
from  anisotropic finite-size scaling for the susceptibility. 
The value reported here $\gamma/\nu_{\perp}\simeq 1.85$  is a 
little bit below the  corresponding RG estimates 
$\gamma/\nu_{\perp}\simeq 2.0$, $\gamma/\nu_{\perp}\simeq 1.98$.
Value $\gamma/\nu_{\perp}\simeq 1.9$ estimated for 
$\theta=1.1$ is in better agreement with theoretical results.
We applied procedure described in this subsection to compute magnetization. 
However it does not give satisfactory results in this case.
This is because magnetization is vanishing quantity at the critical 
point and therefore it is very sensitive to proper determination of 
critical temperature. 

Finally, let us mention that in spite of the substantial computation effort reported in this paper, the numerical values of the critical exponents are not extremely accurate. This is due to the difficulty of the numerical techniques to deal with anisotropic systems 
(see {\it e.g.} Ref.~\onlinecite{Chatelain99}), but from the MC simulations presented in this paper, we believe that we can safely conclude in favor of the anisotropic critical point, since anisotropic scaling is overall nicely confirmed.

\acknowledgments  
This work was supported in part by the~7th FP, IRSES projects No~269139
``Dynamics and Cooperative phenomena in complex physical and biological
environments'' and No~295302 ``Statistical Physics in Diverse Realizations''.

\end{document}